# INDIRECT NEUTRINO OSCILLATIONS

K.S. BABU[a,*], JOGESH C. PATI[b,c,†] and FRANK WILCZEK[b,‡]

[a] *Bartol Research Institute, University of Delaware*

*Newark, DE 19716*

[b] *School of Natural Sciences, Institute for Advanced Study*

*Olden Lane, Princeton, NJ 08540*

[c] *Department of Physics, University of Maryland*

*College Park, MD 20742*

## ABSTRACT

We show how two different scales for oscillations between $e$ and $\mu$ neutrinos, characterized by different mixing angles and effective mass scales, can arise in a simple and theoretically attractive framework. One scale characterizes direct oscillations, which can accommodate the MSW approach to the solar neutrino problem, whereas the other can be considered as arising indirectly, through virtual transitions involving the $\tau$ neutrino with a mass $\sim 1$ eV. This indirect transition allows the possibility of observable $\bar{\nu}_\mu \leftrightarrow \bar{\nu}_e$ oscillations at accelerator and reactor energies. We discuss specifically the parameters suggested by a recent experiment at Los Alamos within this framework.

---

\* Address [b] starting September 1995. Research supported in part by DOE Grant DE-FG02-91ER406267, BABU@BARTOL.UDEL.EDU

† Address [c] after July 1, 1995. Research supported in part by NSF Grant PHY-9421387 and in part by sabbatical leave grant from the University of Maryland, PATI@UMDHEP.UMD.EDU

‡ Research supported in part by DOE Grant DE-FG02-90ER40542, WILCZEK@SNS.IAS.EDU

The dearth of neutrinos observed to be emanating from the sun, compared to theoretical expectations, may be caused by the oscillation of these neutrinos, born as $\nu_e$, into other types having smaller cross-sections at the detector.[1] Both matter-induced (MSW)[2] and vacuum ("just so")[3] oscillations have been invoked in this regard. The range of (mass)$^2$ differences of interest are of order $\sim (10^{-5}-10^{-4})$ eV$^2$ or $\sim (10^{-11}-10^{-10})$ eV$^2$ in the respective cases. These mass scales are considerably smaller than those of primary interest for accelerator oscillation experiments, which are of order $\sim 1$ eV$^2$. Thus if $\nu_e \leftrightarrow \nu_\mu$ oscillations were to be observed in an accelerator experiment, it would appear at first sight as if one were confronted with some rather peculiar alternatives, *e.g.* that the relevant oscillation for the solar neutrino problem is $\nu_e \leftrightarrow \nu_\tau$, and the mass of $\nu_e$ is much closer to that of $\nu_\tau$ than to that of $\nu_\mu$; or that some other hitherto undiscovered neutrino type is involved; or that the solar neutrino problem is solved in some other way than through neutrino oscillations.

In this brief note we shall discuss another alternative, slightly subtler but seemingly quite natural, which is fully consistent with the existence of $\nu_e \leftrightarrow \nu_\mu$ oscillations in both settings (solar and accelerator). It arises from a pattern of neutrino masses and mixings that has been suggested on independent theoretical grounds,[4] as we shall recall below.

## 1. Indirect Mixing

We assume that the electron neutrino $\nu_e$ can be expressed in terms of mass eigenstates $\nu_j$, $j = 1, 2, 3$ in the form

$$\nu_e = \sum_{j=1}^{3} U_{ej} \nu_j \qquad (1.1)$$

and similarly for $\nu_\mu, \nu_\tau$. Possible mixing with heavier neutrinos, if any, will be assumed to be negligible so that the mixing matrix $U$ is unitary. Thus a muon antineutrino emitted at time zero evolves into the superposition

$$\bar{\nu}_\mu(t) = \sum_{j=1}^{3} U^*_{\mu j} \exp(-iE_j t)\bar{\nu}_j \qquad (1.2)$$

at time $t$, where $E_j = \sqrt{m_j^2 + p^2} \approx p + m_j^2/2p$ for the masses and momenta of interest. We wish to consider the possibility that $m_1$ and $m_2$ are very small relative to $m_3$, such that we may ignore them. That is, the phase accumulations $\exp(-im_j^2 L/2p)$ are supposed to differ very little from unity for $j = 1, 2$ and the lengths $L$ and momenta $p$ characteristic of accelerator experiments. This will embody our motivation above, by having $m_1$ and $m_2$ of the magnitude suggested by the solar neutrino problem (or smaller, in that we may have $m_1 \ll m_2$). Under these hypotheses, the probability for oscillations among various species, emitted at energy (or momentum) $E$ to be observed at a distance $L$ is given by

$$\begin{aligned}
|\langle \nu_e(0)|\nu_\tau(L)\rangle|^2 &= 4\sin^2\left(\frac{m_3^2 L}{4E}\right)|U_{e3} U_{\tau 3}|^2 \\
|\langle \nu_\mu(0)|\nu_\tau(L)\rangle|^2 &= 4\sin^2\left(\frac{m_3^2 L}{4E}\right)|U_{\mu 3} U_{\tau 3}|^2 \\
|\langle \bar{\nu}_\mu(0)|\bar{\nu}_e(L)\rangle|^2 &= 4\sin^2\left(\frac{m_3^2 L}{4E}\right)|U_{e3} U_{\mu 3}|^2 \ .
\end{aligned} \qquad (1.3)$$

In the interesting case $1 \simeq |U_{\tau 3}|^2 \gg |U_{e3}|^2, |U_{\mu 3}|^2$, it is natural to say that $\bar{\nu}_\mu \to \bar{\nu}_e$ oscillation proceeds indirectly, through a virtual $\bar{\nu}_\tau$. The existence of such indirect mixing, of course, does not preclude the possibility of oscillations – conceivably much larger in amplitude – that become visible only at larger values of $L/E$, as for the case of solar neutrinos. It provides the slightly subtle alternative to which we previously alluded.

## 2. Numerical Parameters

An appealing feature of (1.3) is that it ties together three different types of oscillations, that is $\bar{\nu}_\mu \to \bar{\nu}_e$, $\bar{\nu}_e \to \bar{\nu}_\tau$, and $\nu_\mu \to \nu_\tau$. Experimental constraints on the latter two processes can be combined to bound the first.

For concreteness, let us assume $m_3^2 = 2$ eV$^2$. From disappearance experiments at the Bugey nuclear reactor one has the bound[5]:

$$|U_{e3}|^2 \leq .02 \ . \tag{2.1}$$

From Fermilab experiment E531 and the CHARM-II experiment, one has the bound[6]:

$$|U_{\mu 3}|^2 \leq .018 \ . \tag{2.2}$$

By combining these, we find the upper bound for indirect mixing:

$$4|U_{e3}|^2 |U_{\mu 3}|^2 \leq 1.5 \times 10^{-3} \ . \tag{2.3}$$

Let us compare this with the results recently reported by Athanassopoulos *et al.*[7] They indicate a $\bar{\nu}_\mu \to \bar{\nu}_e$ oscillation probability of $(3.4 \pm 1.8 \text{ (stat.)} \pm 0.7) \times 10^{-3}$, and the mixing parameter $\sin^2 2\theta$ (deduced from their Fig. 3), which corresponds to the LHS of eq. (2.3), is nearly $(1.2 - 2.5) \times 10^{-3}$, for $\Delta m^2 \simeq 2$ eV$^2$. We see that these results are compatible with the indirect mixing hypothesis, though not by a wide margin. Clearly it would be absurd to claim this in any way as confirmation of the hypothesis; but we feel it does add some additional interest and plausibility to possible mixings of the order of magnitude being explored in the LAMPF experiment.

Bounds for other values of $\Delta m^2$ are indicated in Table 1.

**Table. 1**: Limits on the mixing parameters $|U_{\mu 3}|$ and $|U_{e3}|$ from ref. 6 and 5 respectively as a function of $\Delta m^2 \simeq m_3^2$. In deriving these limits, we have assumed $|U_{\tau 3}| \simeq 1$ and used the unitarity relation $|U_{\tau 3}|^2 + |U_{\mu 3}|^2 + |U_{e3}|^2 = 1$ to determine $|U_{\tau 3}|$ iteratively and in turn $U_{e3}, U_{\mu 3}$ (via eq. (1.3)). The last column corresponds to the expected mixing probability for $\bar{\nu}_\mu \leftrightarrow \bar{\nu}_e$ oscillation at accelerators.

| $\Delta m^2 (\text{eV}^2)$ | $4|U_{\mu 3}|^2_{max}$ | $4|U_{e3}|^2_{max}$ | $(4|U_{\mu 3}|^2 |U_{e3}|^2)_{max}$ |
|---|---|---|---|
| 0.5 | 0.25 | 0.04 | $2.5 \times 10^{-3}$ |
| 1 | 0.09 | 0.06 | $1.4 \times 10^{-3}$ |
| 2 | 0.07 | 0.08 | $1.5 \times 10^{-3}$ |
| 4 | 0.05 | 0.15 | $2.0 \times 10^{-3}$ |
| 6 | 0.03 | 0.17 | $1.1 \times 10^{-3}$ |
| 8 | 0.018 | 0.17 | $0.77 \times 10^{-3}$ |

## 3. Theoretical Context

Patterns of masses and mixing angles that allow accessible rates of indirect $\bar{\nu}_\mu \to \bar{\nu}_e$ oscillation as discussed above, and are compatible both with limits obtained from searches for direct $\bar{\nu}_e \leftrightarrow \bar{\nu}_\tau$ and $\nu_\mu \leftrightarrow \nu_\tau$ oscillations[5,6] and with the MSW solution to the solar neutrino problem are $|U_{e3}| \approx (.10 - .16)$, $|U_{\mu 3}| \approx (.11 - .15)$ for $m_3^2 \approx (.5 - 3)$ eV$^2$; and $m_1 < m_2 \sim (2 - 3) \times 10^{-3}$ eV, $|U_{\mu 1}| \sim |U_{e2}| \approx (2.5 - 5) \times 10^{-2}$ for the "small angle" MSW solution[1], or $m_1 < m_2 \sim (2 - 10) \times 10^{-3}$ eV, $|U_{\mu 1}| \sim |U_{e2}| \approx (.43 - .65)$ for the "large angle" MSW solution. It seems appropriate to mention now that this qualitative pattern of masses, and to a lesser extent of angles, has been suggested on quite independent grounds in the context of theoretical attempts to correlate quark and charged lepton, and predict neutrino, mass parameters.

Light neutrino masses with a hierarchy as exhibited above arise naturally in the context of a large class of unified gauge models – *e.g.* those with left-right symmetric gauge structures, which must include standard model singlet right-handed neutrinos $\nu_R^i$, such as $SO(10)$ or its subgroup $SU(4) \times SU(2) \times SU(2)$. These models realize the famous see–saw mechanism, in which the $\nu_R^i$ acquire large Majorana masses $M_i$. When these are combined with hierarchical Dirac masses $m_{Di}$, one has the see–saw relation $m_i \approx m_{Di}^2/M_i$ for the physical neutrino masses. If we assume that the $M_i$ are all $\sim 10^{12}$ GeV (within a factor of 10 (say)), then one finds the desired qualitative pattern of physical masses for $m_{D1,2,3} \approx 1$ MeV, 300 MeV, 80 GeV, which are quite reasonable orders of magnitude to expect for the scale of Dirac masses in the corresponding families (compare with the masses of the $u, c$ and $t$ quarks). The mass scale $\sim 10^{12}$ GeV for $M_i$ is particularly intriguing because it has been suggested in other contexts, as the scale for Peccei-Quinn symmetry breaking, or for supersymmetry breaking in a hidden sector, or for preon–binding in a SUSY–composite model.

Theoretical ideas regarding mixing angles are even more tentative. One interesting idea[8], that has had some success in providing a simple understanding of the inter-family mass hierarchy $m_{u,d,e} \ll m_{c,s,\mu} \ll m_{t,b,\tau}$ and, with additional hypotheses, other important qualitative aspects of the quark and lepton mass matrices, deserves special mention. According to this idea the Dirac masses of the neutrinos as well as those of quarks and charged leptons arise indirectly through mixing with heavier vector–like families with masses $\sim 1$ TeV (generalized see–saw).[9] If there is just one such vectorial family, which is a doublet of $SU(2)_L$ or $SU(2)_R$, then only one light family receives a mass. With two vectorial families having the quantum numbers of a $\mathbf{16}$ and a $\mathbf{\overline{16}}$ of $SO(10)$, one obtains a hierarchical pattern of light masses for the three light families, and a parameter $p \approx 2\sqrt{m_\mu^0/m_\tau^0} \approx (1/2 \text{ to } 1/3)$, characterizing the $\mu - \tau$ mass hierarchy (at a high scale), appears.[10] For details of a specific model of this type see ref. 4, especially case 2. This specific model suggests not only neutrino masses in the range mentioned above, but also sizable ($\sim 5 - 15\%$) $\nu_\mu - \nu_\tau$ and $\nu_e - \nu_\tau$ mixings, with the relation $U_{e2} \approx U_{\mu 1} \approx (U_{e3})(\frac{2}{p})$. Thus within this model only the large-angle MSW solution is compatible with the hypothesis of indirect $\bar{\nu}_\mu \leftrightarrow \bar{\nu}_e$ oscillation.

To summarize: the suggestion of indirect oscillation presented here raises the interesting possibility that $\bar{\nu}_\mu \leftrightarrow \bar{\nu}_e$ oscillations in accelerator experiments, if observed, could reflect the mass of $\nu_\tau \sim$ (1 to few) eV, allowing $\nu_\tau$ to serve as a cosmologically significant hot component of dark matter; while the depletion of $\nu_e$'s from the sun would reflect direct mixing and $\Delta m^2$ of approximately $10^{-5}$ eV$^2$ for the $\nu_e - \nu_\mu$ system.[11] This scenario requires that not only $\bar{\nu}_\mu \leftrightarrow \bar{\nu}_e$ but also $\nu_\mu \leftrightarrow \nu_\tau$ and $\nu_e \leftrightarrow \nu_\tau$ oscillations occur at levels accessible in the forseeable future.

[10] It needs to be mentioned that within supersymmetric composite models, the coexistence of chiral and vector–like families seems to be a natural feature.[8] Assuming supersymmetry, the number of such vectorial families with masses $\sim 1$ TeV cannot exceed two, regardless of their origin, because otherwise the QCD coupling will grow above 1 TeV and become confining well below $10^{16}$ GeV.

[11] For completeness, we mention that in addition to matter–enhanced direct $\nu_e - \nu_\mu$ transition, the $\nu_e - \nu_\tau$ oscillation suggested here would lead to a further depletion of $\nu_e$ for the solar neutrinos by about $\frac{1}{2}\left(4|U_{e3}|^2\right) \approx (4-5)\%$.